\documentclass[preprint,prc,aps,nofootinbib,showpacs,showkeys]{revtex4}
\usepackage{epsfig}
\usepackage{bm}

\newcommand{\ssst}[1]{\scriptscriptstyle{#1}}
\newcommand{\mN}{m_{\ssst{N}}}

\newcommand{\mtheta}{m_{\ssst{\Theta^+}}}
\newcommand{\rtheta}{R_{\ssst{\Theta^+}}}
\newcommand{\ztheta}{Z_{\ssst{\Theta^+}}}
\newcommand{\btheta}{B_{\ssst{\Theta^+}}}
\newcommand{\thetaplus}{{\Theta^+}}
\begin{document}

\title{Effective mass and decay of $\Theta^+$ in nuclear matter 
in quark-meson coupling model}

\author{C. Y. Ryu$^1$}
\email{cyryu@color.skku.ac.kr}
\author{C. H. Hyun$^{1,2}$}
\author{J. Y. Lee$^{3}$}
\author{S. W. Hong$^{1}$}
\affiliation{
$^1$Department of Physics and Institute of Basic Science, \\
Sungkyunkwan University,
Suwon 440-746, Korea, \\
$^2$School of Physics, Seoul National University,
Seoul 151-742, Korea, \\
$^3$Department of Physics, Sejong University,
Seoul 143-747, Korea}

\date{\today}

\begin{abstract}
The in-medium mass of a $\thetaplus$, $\mtheta^*$, 
in cold symmetric nuclear matter is calculated 
by using the quark-meson coupling model.
The $\Theta^+$ is treated as an MIT bag with the quark content $uudd\bar s$.
Bag parameters for a free $\thetaplus$ are fixed to reproduce
the observed mass of the $\thetaplus$.
In doing so, we use three different values of the $s$-quark mass
since the mass of the $s$-quark is not well known.
As usual, the strengths of the $u$ and $d$ quark couplings to $\sigma$- 
and $\omega$-meson fields are
determined to fit the nuclear saturation properties.
However, the coupling constant $g_\sigma^s$ 
between the $s$-quark and the $\sigma$-meson cannot be fixed
from the saturation properties, 
and thus we treat $g_\sigma^s$ as a free parameter and 
investigate how $\mtheta^*$ depends on $g_\sigma^s$.
%$\mtheta^*$ is calculated up to 2.5 times the nuclear saturation density,
%and we find that 
We find that
$\mtheta^*$ depends significantly on the value of $g_\sigma^s$ 
but not on the mass of the $s$-quark. 
Chemical potentials of the 
$\Theta^+$ and the $K+N$ system are calculated to
discuss the decay of a $\Theta^+$ in nuclear matter.
We calculate the effective mass of a kaon in nuclear matter in two ways;
using the optical potential of $K^-$ in matter and using quark model.
By comparing the effective masses calculated from these two methods,
we find the magnitude of the real part of the optical potential
that is consistent with the usual quark model is about 100 MeV.

\end{abstract}


\maketitle

%\pacs{21.65.+f, 21.80.+a, 12.39.Ba}

\section{Introduction}

Since the mass spectra of pentaquark baryons were first studied with
the MIT bag model \cite{strot} more than two decades ago, 
they were recently reinvestigated in the framework of chiral 
soliton model \cite{dpp97}. 
The experimental mass and decay width of a $\Theta^+$, 
one of the pentaquark baryons,
claimed to be measured by LEPS collaboration \cite{leps03}
seem to agree with the theoretical prediction \cite{dpp97}.
The existence of pentaquark baryons is, however,
still a matter of controversy. There are experiments done 
by different groups showing positive results \cite{diana03,clas03,saphir03},
but some other experiments don't observe the claimed peak \cite{hicks}. 
Though the existence of a $\Theta^+$ remains to be confirmed, in this work 
we assume it exists with the quark content $uudd \bar s$ and 
take the mass $\mtheta$ to be $\approx 1540$ MeV \cite{leps03}.

To probe the pentaquark baryons in matter,
productions of a $\thetaplus$ in relativistic 
heavy ion collisions are investigated theoretically 
at the highest energy of RHIC \cite{jr03}, 
at the lowest SPS energy \cite{letessier03}, and for the transition phases
from quark-gluon plasma to hadronic states \cite{tam03}.
It is believed that the mass of hadrons changes in hot and/or dense systems
due to their interaction with the surrounding matter.
Results from the measurements of dileptonic decays of $\rho$- and $\omega$-
mesons in heavy ion collisions \cite{ceres,kek,trnka05} 
can be interpreted as changes of hadron masses in medium.
One can expect that the mass of a $\thetaplus$ in medium,
$\mtheta^*$, may also be different from its free mass. 
 If such a change in mass occurs indeed, it can influence
the production rate of $\thetaplus$ in heavy ion collision 
experiments.

In this work, as a precursory step to the studies of $\mtheta^*$
in hot and/or dense matter,
we consider a $\thetaplus$ in cold matter
%.
%Even though the states formed in heavy ion 
%collisions are expected to be hot and/or dense,
%consideration of cold matter is a first step 
to fix the parameters of the models.
Recently, $\mtheta^*$ at the nuclear saturation density
and zero temperature has been calculated 
in two independent schemes \cite{kll-04,st-04}, 
whose conclusions are quite different from each other.
In Ref. \cite{kll-04},
$\mtheta^*$ does not nearly change from its free mass, 
but in Ref. \cite{st-04} a considerable change in mass 
is obtained.
In this work we treat the dense matter in terms of
the quark-meson coupling (QMC) model \cite{qmc} and 
calculate the effective mass of a $\Theta^+$ in nuclear matter.
In the QMC model, baryons are treated as 
MIT bags, which interact with each other through the exchange of mesons 
such as $\sigma$ and $\omega$.
Consistently with the QMC model,
we also treat the $\thetaplus$ as an MIT bag.
Then additional three bag parameters for a $\Theta^+$ need to be fixed; 
bag radius $\rtheta$, bag constant $\btheta$ and a 
phenomenological constant $\ztheta$. 
We take $\rtheta$ values to be 0.6, 0.8 and 1.0 fm.
$\btheta$ and $\ztheta$
are fixed to reproduce the free mass $\mtheta$ for given $\rtheta$ values. 
For the interactions between quarks ($u$, $d$ and $s$)
and mesons ($\sigma$ and $\omega$), 
four quark-meson coupling constants
$g_\sigma^{u(d)}$, $g_\sigma^s$, $g_\omega^{u(d)}$ and $g_\omega^s$
are introduced. 
(We assume 
$g_\sigma^u = g_\sigma^d$ and $g_\omega^u= g_\omega^d$.) 
$g_\sigma^u$ and $g_\omega^u$ are fitted as usual to reproduce
the binding energy per nucleon $E_b$ (= 16.0 MeV) at 
the nuclear saturation density $\rho_0$ ($=0.17\, {\rm fm}^{-3}$).
We also use the so-called modified QMC (MQMC) model \cite{mqmc},
in which the bag constant $B$ is assumed to depend on density.
These nuclear models will be presented in Section \ref{sec:matter}.

In quark models, $g^s_\sigma$ and $g^s_\omega$ are often 
set to zero. %following the OZI rule [{\bf REF}].
However, studies of hypernuclei \cite{shen02, yfuji} or
kaon-nucleon systems \cite{gal, oset} 
seem to indicate that
the interaction between the $s$-quark and $u\, (d)$ quarks could be
non-negligible.
Recently, deeply bound kaonic states are
predicted theoretically for light systems \cite{ay-prc02,dote04},
and indeed KEK-E471 has revealed similar states identified as
deeply bound tribaryon-kaon systems \cite{suzuki-plb04,suzuki-05}.
%{\bf Put arguments related with optical potential.}
In some cases \cite{shen02, yfuji, oset}, 
an interaction weaker than what is expected by quark models is
favored, but in some other cases \cite{gal, ay-prc02,dote04} 
a strong interaction seems to be needed.
Due to this large uncertainty in the interaction of strange sector,
we treat $g^s_\sigma$ as a free parameter and study the dependence of 
$\mtheta^*$ on $g_\sigma^s$.
However, for the interaction between the $s$-quark and 
the $\omega$-meson, we assume the quark model value $g^s_\omega = 0$.

%In the calculations of $\mtheta^*$, 
%we also use the MQMC model as well as QMC model for $\thetaplus$.
%These QMC and MQMC models for $\Theta^+$ need to be distinguished from
%those for nuclear matter. 
%The mass of the exchanged mesons can also be affected by
%the matter effect.
%We take into account such an effect in the QMC and the MQMC models
%by treating
%the $\omega$-meson as a bag consisting of a quark and an anti-quark,
%which couple with the $\sigma$-meson.
%These models are referred to as QMC-MB and MQMC-MB, respectively,
%MB meaning a meson bag.
%Bag parameters for the $\omega$-meson are also determined 
%to reproduce its free mass $m_\omega =783$ MeV.
%On the other hand, we assume the $\sigma$-meson as a point particle
%with a conventional value of mass, $m_\sigma = 550$ MeV.
Stability of the $\Theta^+$ particle in medium may be discussed by
comparing the chemical potential of a $\Theta^+$ with that of a $KN$ system,
which is a possible decay channel of $\Theta^+$. 
If the chemical potential
of a $\Theta^+$ is lower than that of a $KN$ system, one cannot
exclude the possibility of a stable $\Theta^+$ in nuclear matter.
In evaluating the chemical potential of $K$, we first need to 
calculate the effective mass of $K$, $m_K^*$.
In subsection III B we show how we calculate $m_K^*$ in two
different methods and estimate the magnitude of the optical potential of $K^-$,
which is under debate by different authors 
\cite{shen02, yfuji,gal, oset,ay-prc02,dote04}.
%In some recent works \cite{st-04,cabrera04,nagahiro04,oset05},
%$\Theta^+$ energy in finite nuclei is
%considered, and it is shown that the existence 
%of $\Theta^+$ hypernuclei depends 
%on the $\Theta^+$ potential in matter.

In Sect.~\ref{sec:model} , we determine the
bag parameters of the nucleon and a $\thetaplus$
from their free masses. The coupling constants between
$u\, (d)$ quarks and mesons are
adjusted to reproduce the known saturation properties. 
In Sect.~\ref{sec:result} , we show the results of $\mtheta^*$
from QMC and MQMC models.
Dependence of the effective mass $\mtheta^*$
on the coupling constant $g_\sigma^s$
between the $s$-quark and $\sigma$ is presented. 
The in-medium decay of a $\Theta^+$ to kaon-nucleon state is 
discussed from the consideration of chemical potentials of a $\Theta^+$
and a $KN$ system.
A summary is given in Sect.~\ref{sec:summary}.

\section{Models \label{sec:model}}
\subsection{QMC and MQMC models for
nuclear matter \label{sec:matter}}

In the QMC model, the nucleons in nuclear matter 
are assumed to be described by static MIT bags in which 
quarks couple to effective meson fields,
which are treated as classical in a mean field approximation. The quark
field $\psi_q$ inside the bag then satisfies the Dirac equation
\begin{eqnarray}
\left[ i \gamma \cdot \partial - ( m_q - g^q_\sigma\, \sigma) 
- g^q_\omega\, \gamma^0 \, \omega_0 \right]\, \psi_q = 0,
\end{eqnarray}
where $m_q$ is the bare mass of the quark and
$\sigma$ and $\omega_0$ are the mean fields of $\sigma$- and
$\omega$-mesons, respectively.
We assume $m_q = 0$ for $q=u$ and $d$ and treat the strange quark mass
$m_s$ as a free parameter.
The ground state solution to the Dirac equation is given by
\begin{eqnarray}
\psi(\bm{r},\, t) = 
{\cal N}_q \exp(-i \epsilon_q t/ R) 
\left(
\begin{array}{c}
j_0(x_q\, r/R) \\
i\, \beta_q\, \bm{\sigma}\cdot\hat{\bm{r}}\, j_1(x_q\, r/R)
\end{array}
\right)
\frac{\chi_q}{\sqrt{4 \pi}},
\end{eqnarray}
where ${\cal N}_q$ is the normalization factor, $R$ is the bag radius and
\begin{eqnarray}
\epsilon_q &=& \Omega_q + R\, g^q_\omega\, \omega_0, \\
%(\epsilon_{\bar q} &=& \Omega_q - R\, g^q_\omega\, \omega_0,) \\
\beta_q &=& \sqrt{\frac{\Omega_q - R\, m^*_q}{\Omega_q + R\, m^*_q}}, \\
\Omega_q &=& \sqrt{x^2_q + (R\, m^*_q)^2}, \\
m^*_q &=& m_q - g^q_\sigma\, \sigma.
\end{eqnarray}
$\chi_q$ is the quark spinor and $x_q$ is determined from the
boundary condition on the bag surface
\begin{eqnarray}
j_0(x_q) = \beta_q\, j_1(x_q).
\label{eq:boundcond}
\end{eqnarray}
The energy of the nucleon bag with the ground state quarks is
given by 
\begin{eqnarray}
E_N &=& \sum_q  \frac{\Omega_q}{R} - \frac{Z_N}{R}
+ \frac{4}{3} \pi\, R^3\, B_N,
\label{eq:bagery}
\end{eqnarray}
where $B_N$ and $Z_N$ are the bag constant and 
a phenomenological constant for the zero-point motion of the nucleon,
 respectively.
The mass of a free nucleon (or the effective mass of a nucleon in matter)
is given by \cite{fbsy90}
\begin{eqnarray}
m^*_N = \sqrt{E^2_N - \sum_q  \left(\frac{x_q}{R} \right)^2}.
\label{eq:efmass}
\end{eqnarray}
We consider three values of $R_0$ ($=$ 0.6, 0.8 and 1.0 fm)
as the bag radius of a free nucleon.
For each $R_0$ value, $B_N$ and $Z_N$ can be 
determined from the minimum condition 
\begin{eqnarray}
\frac{\partial m^*_N}{\partial R} = 0
\label{eq:minimum}
\end{eqnarray}
by taking $m_N^*$ as $\mN$ = 939 MeV.
The values of $B_N^{1/4}$ and $Z_N$ are determined to be
(188.1 MeV, 2.03), (157.5 MeV, 1.628) and (136.3 MeV, 1.153)
for $R_0$=0.6, 0.8, and 1.0 fm, respectively.
When Eq.~(\ref{eq:minimum}) is applied for a nucleon in matter,
we obtain the effective bag radius $R$ as well as 
the effective mass of nucleons $m^*_N$.

The self-consistency condition (SCC) for $\sigma$-meson fields
is obtained through the thermodynamic condition
\begin{eqnarray}
\frac{\partial \varepsilon}{\partial \sigma} = 0
\label{eq:scc}
\end{eqnarray}
with the energy density of the matter
\begin{equation}
\varepsilon = \frac{1}{2} m^2_\sigma \sigma^2 + 
\frac{1}{2} {m_\omega}^2 \omega^2_0 + 
4 \int^{k_F}_0 \frac{d^3 k}{(2 \pi)^3} \sqrt{k^2 + \mN^{*2}},
\label{eq:enden}
\end{equation}
where $k_F$ is the Fermi momentum of the nucleons at a given density.
Then the effective mass $m_N^*$ can be calculated self-consistently
by solving the boundary condition of Eq.~(\ref{eq:boundcond}),
the minimum condition of Eq. (\ref{eq:minimum}), and
the SCC of Eq.~(\ref{eq:scc}). 
The quark-meson coupling constants $g_\sigma^u$ and $g_\omega^u$
are fixed to reproduce
the saturation properties; $\rho_0$ and $E_b$.
The coupling constants thus fixed, the resulting $m_N^*$, 
and the compression modulus $K$ from the QMC model
are listed in Table \ref{tab:qmc-mqmc}.

\begin{table}[t]
\begin{center}
\begin{tabular}{|c|c|c|c|c||c|c|c|c|c|}\hline
 & \multicolumn{4}{c||}{QMC} &
 \multicolumn{5}{c|}{MQMC} \\ \cline{2-10}
$R_0$(fm) & $g^u_\sigma$ & $g^u_\omega$ & $\mN^*/\mN$ & $K$ (MeV) 
&$g_\sigma^u$ & $g^u_\omega$  & $g^N_\sigma$ & $\mN^*/\mN$ & $K$ (MeV) \\ \hline
0.6 & 5.307 & 1.476 & 0.89 & 223.1
& 1.0 & 2.705 & 6.805 & 0.78 & 292.2 \\ \hline 
0.8 & 5.516 & 1.256 & 0.91 & 200.7 
& 1.0 & 2.676 & 6.823 & 0.79 & 287.7 \\ \hline
1.0 & 5.571 & 1.150 & 0.91 & 190.1
& 1.0 & 2.665 & 6.825 & 0.79 & 287.3 \\ \hline
%1.2 & 5.561 & 1.090 & 185.1 & 0.92 
%& 6.805 & 2.655 & 287.7 & 0.79 \\ \hline
\end{tabular}
\end{center}
\caption{The coupling constants, the effective mass $m_N^*$
and the compression modulus $K$ at the saturation density
for the QMC and the MQMC models.}
\label{tab:qmc-mqmc}
\end{table}

The MQMC model for nuclear matter takes into account
the density dependence of $B_N$, which may be expressed as \cite{mqmc}
\begin{eqnarray}
B_N (\sigma) = B_N \left( 1 - g^N_\sigma\, \frac{4}{\delta} 
\frac{\sigma}{m_N} \right)^\delta.
\label{eq:bagcons}
\end{eqnarray}
We take the limit $\delta \rightarrow \infty$, in which case 
\begin{eqnarray}
B_N (\sigma) = B_{N} \exp(-4\, g^N_\sigma \sigma / m_N).
\label{eq:mqmcbag}
\end{eqnarray}
In this model we have an extra parameter $g_\sigma^N$ to be determined. 
We use the following two constraints at the saturation to fix $g^N_\sigma$.
\[
\mN^* = (0.7 - 0.8) \mN,\,\,\,\,
K = (200 - 300)\, \mbox{MeV}.
\]
For all $R_0$ values, 
we have used a fixed value of $g_\sigma^u =1$ \cite{mqmc} in this work.

The self-consistency condition for the MQMC model is also obtained 
from Eq.~(\ref{eq:scc}). It has additional terms due to 
the density dependent bag constant \cite{mqmc}.
The coupling constants $g_\sigma^N$ and $g_\omega^u$ and the resulting
$m_N^*$ and $K$ in the MQMC model are also
listed in Table \ref{tab:qmc-mqmc}.

\subsection{QMC and MQMC models for $\Theta^+$ in medium}

We may regard a $\thetaplus$ in medium also as an MIT bag,
and write its effective mass as
\begin{eqnarray}
m_{\Theta^+}^* = \sqrt{E_{\Theta^+}^2 - \sum_{i=q, \bar q} 
\left( \frac{x_i}{R} \right)^2}, 
\label{eq:mtheta}
\end{eqnarray}
where the bag energy of a $\Theta^+$ is given by
\begin{eqnarray}
E_{\Theta^+} &=& \sum_{i=q, \bar q} \frac{\Omega_i}{R}
- \frac{Z_{\Theta^+}}{R} + \frac 43 \pi R^3 B_{\Theta^+} 
\label{eq:etheta} \\
\Omega_i &=& \sqrt{x_i^2 + (Rm_i^*)^2} \label{eq:Omegatheta} \\
m_i^* &=& m_i - g_\sigma^i \sigma, \label{eq:mqeffective}
\end{eqnarray}
in which $i$ includes $\bar s$ as well as $u$ and $d$ quarks.
For fixed $R_0$ values, $B_{\Theta^+}$ and $Z_{\Theta^+}$ 
can be determined from the observed value of 
the free $\Theta^+$ mass ($m_{\Theta^+}$ = 1540 MeV)
\cite{leps03} and a minimum condition similar 
to Eq. (\ref{eq:minimum}) but for $\Theta^+$. 
Since the value of $m_s$ is not well known, 
we extract
$B_{\Theta^+}$ and $Z_{\Theta^+}$ for a few different 
values of $m_s$ and $R_0$.
The results are shown in Table \ref{tab:zb}.
After fixing the bag parameters for $\Theta^+$ as above, 
we calculate its effective mass $\mtheta^*$ in nuclear matter
by using Eq.~(\ref{eq:mtheta}). 
$\mtheta^*$ deviates from $m_{\Theta^+} = 1540$ MeV
since $\sigma$ in Eq.~(\ref{eq:mqeffective}) is non-zero.

\begin{table}
\begin{center}
\begin{tabular}{|c|c|c|c|c|c|c|}\hline
% &\multicolumn{2}{c|}{  }&
&\multicolumn{2}{c|}{ $m_s$ = 50 MeV}&
\multicolumn{2}{c|}{ $m_s$ = 150 MeV} &
\multicolumn{2}{c|}{ $m_s$ = 300 MeV} \\ \cline{2-7}
 $R_0$(fm)  & 
%$\bN^{1/4}$ & $\zN$ &
$\btheta^{1/4}$ & $\ztheta$ &
$\btheta^{1/4}$ & $\ztheta$ &
$\btheta^{1/4}$ & $\ztheta$ \\ \hline
0.6 &
%188.1 & 2.030 &
219.6 & 4.557 &
217.9 & 4.653 &
213.4 & 4.804 \\ \hline
0.8 &
%157.5 & 1.628 &
182.3 & 3.805 &
180.2 & 3.941 &
176.8 & 4.158  \\ \hline
1.0 & 
%136.3 & 1.153 &
156.8 & 2.943 &
155.4 & 3.122 &
151.8 & 3.413  \\ \hline
%1.2 & 120.6 & 0.624 & 
%138.4 & 1.899 &
%136.8 & 2.233 &
%133.5 & 2.603 \\ \hline
\end{tabular}
\end{center}
\caption{$B_{\Theta^+}$ and $Z_{\Theta^+}$ values for a $\Theta^+$ when
$R_0$ = 0.6, 0.8 and 1.0 fm and $m_s$ = 50, 150 and 300 MeV.
$B^{1/4}_{\Theta^+}$ is in MeV. In all the cases, $m_u = m_d = 0$.}
\label{tab:zb}
\end{table}

If we consider the case when the bag constant $B_{\Theta^+}$ 
may depend on the matter density,
we refer to it as the MQMC model for a $\Theta^+$.
Using the form of Eq.~(\ref{eq:mqmcbag}), we express $B_{\Theta^+}$ as
\begin{eqnarray}
B_{\Theta^+}(\sigma) = B_{\Theta^+} ~ 
{\rm exp}(-4 {g'}_\sigma^B \sum_{q=u,d} n_q \sigma / m_{\Theta^+})
\label{eq:thetabag}
\end{eqnarray}
where $\sum n_q$=4 for a $\Theta^+$ bag and
the extra parameter ${g'}_\sigma^B$ can be 
related to $g_\sigma^N$ of Eq.~(\ref{eq:mqmcbag})
by ${g'}_\sigma^B$=$g_\sigma^N$/3 \cite{pal}
with $g^N_\sigma$ given
in Table \ref{tab:qmc-mqmc}.

As noted in the Introduction, the nuclear matter properties 
at the saturation can fix the parameters related to
$u$- and $d$-quarks, but not $s$-quarks.
We thus treat $g_\sigma^s$ as a free parameter in the following 
and show our results for three choices of $g_\sigma^s$ values.

\section{Results \label{sec:result}}
\subsection{Effective mass of $\Theta^+$}

%%%%%%%%%%%%%%%%%%%%%%%%%%%%%%%%%%%%%%%%%%%%%%%%%%%%%%%%%%%%%%%%%%%%%%%%%%%%%%
\begin{table}[tbp]
\begin{center}
\begin{tabular}{|c|c|c|c||c|c|c|}\hline
 & \multicolumn{3}{c||}{$m_{\Theta^+}^* / m_{\Theta^+}$ in QMC} &
   \multicolumn{3}{c|}{$m_{\Theta^+}^* / m_{\Theta^+}$ in MQMC} \\ \cline{2-7}
   $R_0$(fm) & $m_s=50$ & $m_s=150$ & $m_s=300$
   & $m_s=50$ & $m_s=150$ & $m_s=300$
   \\ \hline
   0.6 &  0.90 & 0.90 & 0.90
   & 0.82 & 0.83 & 0.84 \\ \hline
   0.8 &  0.91 & 0.91 & 0.91
   & 0.82 & 0.83 & 0.84 \\ \hline
   1.0 &  0.92 & 0.92 & 0.92
   & 0.82 & 0.83 & 0.84 \\ \hline
%   1.2 &  0.92 & 0.92 & 0.92 
%   & 0.85 & 0.86 & 0.87 \\ \hline\hline
%    & \multicolumn{3}{c||}{$m_{\Theta^+}^* / m_{\Theta^+}$ in QMC-MB} &
%       \multicolumn{3}{c|}{$m_{\Theta^+}^* / m_{\Theta^+}$ in MQMC-MB} \\ \cline{2-7}
%       $R_0$(fm) & $m_s=50$ & $m_s=150$ & $m_s=300$
%       & $m_s=50$ & $m_s=150$ & $m_s=300$
%      \\ \hline
%       0.6 & 0.90 & 0.90 & 0.90
%           & 0.88 & 0.88 & 0.89 \\ \hline
%       0.8 & 0.91 & 0.91 & 0.91
%           & 0.88 & 0.88 & 0.89 \\ \hline
%       1.0 & 0.91 & 0.91 & 0.91
%           & 0.88 & 0.88 & 0.89 \\ \hline
\end{tabular}
\end{center}
\caption{$m_{\Theta^+}^* / m_{\Theta^+}$ at the saturation density
for three different
bare masses of the strange quark.
The table shows $m_{\Theta^+}^* / m_{\Theta^+}$ is
insensitive to the choice of $m_s$ and $R_0$. $g_\sigma^s = 0$
is used for the calculations shown in this table.
$m_s$ is in MeV.}
\label{tab:redumas}
\end{table}
%%%%%%%%%%%%%%%%%%%%%%%%%%%%%%%%%%%%%%%%%%%%%%%%%%%%%%%%%%%%%%%%%%%%%%

\begin{figure}
\begin{center}
\epsfig{file=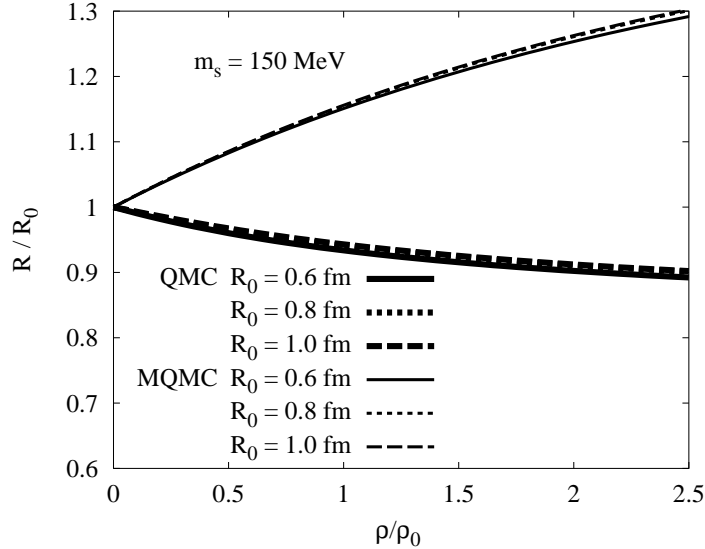,height=7.5cm,angle=0}
\end{center}
\caption{Ratios of the effective radius $R$ of a $\Theta^+$ bag 
to $R_0$ for $m_s = 150$ MeV. $R/R_0$ from QMC(MQMC) calculations are
plotted by thick(thin) curves. $g_\sigma^s = 0$ is used for
calculations in this figure.}
\label{fig:bagrad}
\end{figure}

Table~\ref{tab:redumas} shows $\mtheta^*$ at the saturation density 
divided by its free mass $\mtheta$ for different bag radii $R_0$
and $s$-quark masses $m_s$. 
$m_{\Theta^+}^* / m_{\Theta^+}$ values are rather stable against
the variation of $R_0$ and $m_s$. 
In the present models, a change in mass is caused by the 
non-zero value of the $\sigma$-field.
Table \ref{tab:qmc-mqmc} shows that $m_N^* / m_N$ at the 
saturation density turns out to be more or less independent of $R_0$, 
which reflects the fact that the value of $\sigma$-field 
is nearly independent of $R_0$ in both QMC and MQMC models.
The effective bag radius $R$ in matter deviates from $R_0$ 
and $x_q$ also changes from that of a free bag due to 
non-zero value of the $\sigma$-field. However, 
as Fig.~\ref{fig:bagrad} shows, the ratio $R/R_0$ depends little
on the choice of $R_0$. 
We also find that the change of $x_q$ value in medium is only
a few \% from its value for a free bag. Since these three
density-dependent quantities, $\sigma$-field, 
$R/R_0$, and $x_q$ are rather independent of 
$R_0$ and $m_s$, $\mtheta^*$ remains more or less the same 
regardless of the values of $R_0$ and $m_s$ for both QMC and MQMC models.

%
%%%%%%%%%%%%%%%%%%%%%%%%%%%% Fig. 1 %%%%%%%%%%%%%%%%%%%%%%%%%%%%%%%%%%%
%%%%%%%%%%%%%%%%%%%%%%%%%%%%%%%%%%%%%%%%%%%%%%%%%%%%%%%%%%%%%%%%%%%%%%%
%
Fig.~\ref{fig:bagrad} also shows that the effective radii of $\Theta^+$
from MQMC (QMC) models increase (decrease) with density of matter.
The increase of bag radii for MQMC models 
is of course due to the change of the bag constant in medium.
The radius increases roughly by $ 15 \%$ at the saturation density,
which may be compared with the results from
previous studies for nucleons \cite{siegel,brown,mardor}.

%%%%%%%%%%%%%%%%%%%%%%%%%%% Fig. 2 %%%%%%%%%%%%%%%%%%%%%%%%%%%%%%%%%%%%%%%%
\begin{figure}[tbp]
\begin{center}
\epsfig{file=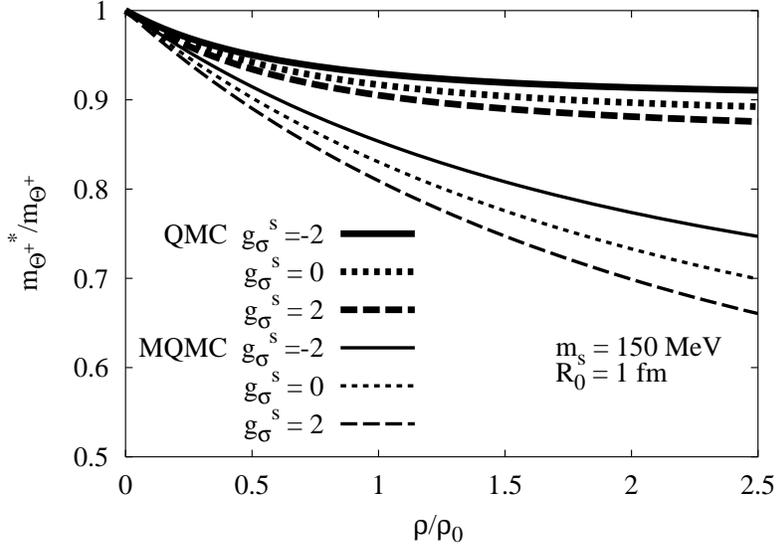,height=7.5cm,angle=0}
\end{center}
\caption{$\mtheta^* / \mtheta$ is plotted
against density for $g_\sigma^s = -2$, 0, 2.
Thick (thin) curves denote $\mtheta^* / \mtheta$ calculated from QMC (MQMC) 
for both $\Theta^+$ and nuclear matter.
}
\label{fig:qmc}
\end{figure}
%%%%%%%%%%%%%%%%%%%%%%%%%%%%%%%%%%%%%%%%%%%%%%%%%%%%%%%%%%%%%%%%%%%%%%%%%%%
In quark models, the $s$-quark -- meson coupling constants are 
usually set to zero,
which leads to the relation between the coupling constants,
$g_{M Y} = \frac{2}{3} g_{M N}$ where $g_{M Y}$ is the meson-hyperon
coupling constant and $g_{M N}$ the meson-nucleon constant.
However, as mentioned in the Introduction, there are studies 
\cite{shen02,yfuji,gal,oset}
indicating that $g_{MY}$ may have a value different
from $\frac{2}{3} g_{MN}$, which means
$g^s_\sigma$ or $ g^s_\omega$ can be non-zero.
In view of this uncertainty in the strange sector,
we choose $g^s_\sigma$ as $-2$, 0 and 2 
and calculate $m_{\Theta^+}^*$ for these choices. 
Some consequences of the non-zero $g^s_\sigma$ value
will be discussed in the next subsection
in connection with the optical potential of a kaon
and its effective mass in nuclear medium.

Fig.~\ref{fig:qmc} shows $\mtheta^* / \mtheta$ calculated 
for $m_s = 150$ MeV and $R_0$ = 1 fm 
at nuclear densities up to $2.5 \rho_0$. 
(As shown in Table \ref{tab:redumas}, other choices of $m_s$ or $R_0$ 
values give us similar results of $m^*_\thetaplus$.)
The solid, dashed and dotted lines correspond to the results with
$g^s_\sigma = -2$, 0 and 2, respectively.
The upper three thick lines represent the results from QMC,
and the lower three thin lines are those from MQMC. 
The biggest change in the mass from the QMC model is about 
12 \% in the region of densities we consider.
$m^*_\thetaplus$ from the QMC model does not change much with
density and $g_\sigma^s$.
This weak dependence on the density and $g_\sigma^s$
is due to small magnitudes of the $\sigma$-field. 
However, this small value of $\sigma$-fields cannot 
account for the spin-orbit phenomenology in nuclei.
The MQMC model \cite{mqmc}
increases the value of $\sigma$-fields 
as much large as in quantum hadrodynamics \cite{qhd}. 
Due to the large $\sigma$-fields, 
$m^*_{\Theta^+}$ from the MQMC model drops rapidly 
with density, and the dependence of 
$m^*_{\Theta^+}$ on the $g_\sigma^s$ value is pronounced.
Another reason for the mass reduction in MQMC is the following. 
As the $\sigma$-field increases with density, 
the bag constant in Eq.~(\ref{eq:thetabag}) decreases.
Since the bag constant term contributes positively to the effective
mass, a smaller bag constant will result in a smaller
effective mass.

\subsection{Kaonic decay of $\Theta^+$ in medium}

Let us consider the in-medium decay of a $\Theta^+$ into
$KN$ channels ($K^+n$ and $K^0p$). Since $K^+ (u \bar s)$ and
its isospin partner $K^0 (d \bar s)$ interact identically 
with symmetric nuclear matter,
we do not distinguish the two decay channels.
If a $\Theta^+$ at rest in free space decays into $KN$,
the momenta of $K$ and $N$
are $|\vec p_K| = |\vec p_N| \approx$ 270 MeV/c.
If the masses and energies of $K, N$ and $\Theta^+$ 
do not change in medium from those in free space,
a $\Theta^+$ in matter at the saturation density
may decay into $KN$
because the momentum of an outgoing $N$ is above the Fermi momentum.
However, since the properties of $K, N$ and $\Theta^+$ may change,
the chemical potentials of these particles in medium need 
to be considered.
%To check whether $\Theta^+$ in medium can decay into $KN$ channels or not,
%we calculate the chemical potentials of $K$, $N$ and $\Theta^+$ in medium.
If the chemical potential of a $\Theta^+$ is larger than the sum of 
those of $K$ and $N$, a $\Theta^+$ in medium can decay into $KN$ channels.
In this subsection, we use only the MQMC model in treating both 
nuclear matter and $\Theta^+$.

In calculating the chemical potential and the effective mass
of a kaon, we can treat a kaon not only as a point particle but also 
as a meson bag. 
By doing that, we may associate the optical potential of a kaon
with $g_{\sigma K}$, the coupling constant between a kaon 
and a $\sigma$-meson, as we will explain shortly.
Among the mean field Lagrangians \cite{kaon1,glend}
suggested for the in-medium kaons,
we employ the Lagrangian given by Ref. \cite{glend}
\begin{eqnarray}
{\mathcal L}_K = D_\mu^* K^* D^\mu K - {m_K^*}^2 K^* K,
\label{eq:kaon}
\end{eqnarray} 
where $K$ denotes the isospin doublet kaon field. The covariant derivative
in the symmetric matter
$D_\mu = \partial_\mu + i g_{\omega K} \omega_\mu$ couples the kaon field
to the $\omega$-meson fields, and 
$g_{\omega K}$ is assumed to follow the quark counting rules, i.e.,
$g_{\omega K} = g_\omega^u$.
The kaon effective mass $m_K^*$
includes the interaction of $K$ with the $\sigma$-meson
and is calculated in the following two methods.

\begin{figure}
\begin{center}
\epsfig{file=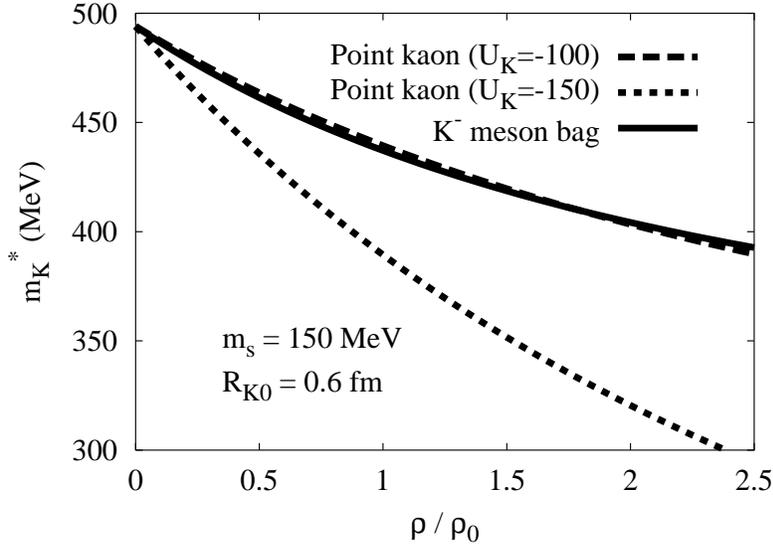,height=7.5cm,angle=0}
\end{center}
\caption{The effective masses of a kaon as a point particle and a meson bag
are plotted by the solid and the dashed curves, respectively. 
The dashed and dotted curves denotes $m_K^*$ when we choose
$U_{\bar K}(\rho_0)$ as $-100$ and $-150$ MeV, respectively.}
\label{fig:kaonmass}
\end{figure}

First, we may treat a kaon as a meson bag. Then the effective
mass of a kaon can be determined from the following equations:
\begin{eqnarray}
m^*_K &=& \sqrt{E^2_K - \sum_{ i = q,\bar{q}} \left( \frac{x_i}{R} \right)^2 },
\label{eq:mK}\\
E_K &=& \sum_{i = q,\bar{q}} \frac{\Omega_i}{R} - \frac{Z_K}{R} +
\frac{4}{3} \pi\, R^3\, B_K.
\label{eq:eK}
\end{eqnarray} 
Here we assume $g_\sigma^s = 0$ and choose
the bag radius of a kaon as
$R_{K0}$ = 0.6 fm and the $s$-quark mass as $m_s$ = 150 MeV.
(The results for other $g_\sigma^s$ values may be inferred 
from Fig. \ref{fig:qmc}.)
By requiring the mass of a free $K^+$ to be $m_K = 494$ MeV, we get 
$B_K^{1/4}$ = 141.65 MeV and $Z_K$ = 1.086.
For a kaon in matter, the bag constant for the kaon 
may be assumed to be
\begin{eqnarray}
B_K (\sigma) = 
B_K \exp\left(- \frac43 g_\sigma^N \frac{\sigma}{m_K} \right),
\label{eq:kaonbag}
\end{eqnarray}
where the factor 3 in denominator is from the quark counting rule.
Using the same MQMC model parameters for nuclear matter
as given in Table \ref{tab:qmc-mqmc},
we obtain $m_K^*$ plotted by the solid curve 
in Fig.~\ref{fig:kaonmass}.
%%%%%%%%%%%%%%%%%%%%%%%%%%% Fig.3  %%%%%%%%%%%%%%%%%%%%%%%%%%%%%%%%%%%%%
%%%%%%%%%%%%%%%%%%%%%%%%%%%%%%%%%%%%%%%%%%%%%%%%%%%%%%%%%%%%%%%%%%%%%%%%

Let us now treat a kaon as a point particle. Then the effective mass
is given as 
\begin{equation}
m_K^* = m_K - g_{\sigma K}\sigma.
\label{eq:pointkaonmass}
\end{equation}
To fix $g_{\sigma K}$, the optical potential for an antikaon ($K^-$) 
in medium is employed.
Note that $g_{\sigma K}$ for $K^+$ is the same as that for $K^-$
as can be seen from ${\mathcal L}_K$ in Eq. (\ref{eq:kaon}).
The real part of the optical potential for antikaons 
at the saturation density %is known to be 
%50 MeV $\lesssim |U_{\bar K}(\rho_0)| \lesssim$ 120 MeV \cite{gal,oset} 
%where 
$U_{\bar K}(\rho_0)$ is given by 
\begin{equation}
U_{\bar K}(\rho_0) = 
- g_{\sigma K} \sigma(\rho_0) - g_{\omega K} \omega_0(\rho_0)
\label{eq:opticalpot}
\end{equation}
in our model.
With $g_{\omega K} = g^u_\omega$ (See Table \ref{tab:qmc-mqmc}.),
we may fix $g_{\sigma K}$ so that $U_{\bar{K}}(\rho_0)$ calculated by
Eq.~(\ref{eq:opticalpot}) becomes a certain value that we choose.
As mentioned in the Introduction, the value of the real part of 
$K^-$ optical potential in matter
is a matter of debate in relation to the existence of a deeply bound
kaonic nuclear system.
We choose $U_{\bar K}(\rho_0) = -100$ MeV because 
this value makes 
$m^*_K$ from Eq.~(\ref{eq:pointkaonmass})
agree with $m^*_K$ of Eq. (\ref{eq:mK}), 
which treats a kaon as a meson bag.
$U_{\bar K}(\rho_0) = -100$ MeV
gives us $g_{\sigma K}$ = 1.997 from Eq. (\ref{eq:opticalpot}).

In Fig.~\ref{fig:kaonmass}, $m^*_K$ obtained for a 
point kaon with $g_{\sigma K} = 1.997$ 
is shown by the dashed curve.
The dashed curve almost overlaps with the solid curve, 
which represents $m^*_K$ for a kaon bag with $g^s_\sigma=0$.
Roughly speaking, the optical potential value 
$U_{\bar{K}}(\rho_0) = -100$ MeV
corresponds to the $g^s_\sigma$ and $g^s_\omega$ values given by the quark model.
A larger magnitude of $U_{\bar{K}}(\rho_0)$ value corresponds to
a larger $g^s_\sigma$ value, and vice versa.
Fig.~\ref{fig:kaonmass} also shows by the dotted curve
$m_K^*$ of a point particle kaon
when we choose $U_{\bar K} = - 150$ MeV. 
$m_K^*$ for $U_{\bar K} = -150$ MeV at the saturation 
is smaller than $m_K^*$ for
$U_{\bar K}(\rho_0) = -100$ MeV by about 10 \%,
and the difference gets larger as the density increases. 
This shows $m_K^*$ is sensitive to the value
of the optical potential at the saturation.
We find that the $g_\sigma^s$ value that gives us
$m_K^*$ nearly overlapping with the dotted curve is
$g_\sigma^s = 2.6$.
$m_K^*$ plotted by the dotted curve is expected to get
extremely small at higher densities, and
may cause kaon condensation in a neutron star. 
Such phenomena will be discussed elsewhere.

%Similarly,
%$m_K^*$ of a kaon as a meson bag is sensitive 
%to $g_\sigma^s$ value as 
%already observed in the behavior of $m_{\Theta^+}^*$ in 
%Fig.~\ref{fig:qmc}. 
%In either case, the effective mass
%of strange hadrons has a non-negligible uncertainty of 
%$\sim$ 10 \%, which can be significant
%and needs to be explored in the future.
Let us now compare the chemical potential of a $\Theta^+$, 
$\mu_{\Theta^+}$, with $\mu_K + \mu_N$ of the $KN$ system.
The equation of motion for the kaon can be written as
$[D_\mu D^\mu+{m_K^*}^2]K=0$.
Using a plane wave for the kaon field,
one obtains the following
dispersion relation with the upper (lower) sign for kaons (antikaons)
in uniform and symmetric matter
\begin{eqnarray}
\epsilon_{K,\bar K} = \sqrt{k_K^2+{m_K^*}^2} \pm g_{\omega K} \omega_0,
\label{eq:chekaon}
\end{eqnarray}
where $k_K$ is the momentum of the kaon. 
The distortion of the kaon field in matter will smear $\epsilon_K$
around this value.
We can then calculate the chemical potential of kaons as
$\mu_K = \epsilon_K$.
The chemical potential of a nucleon and a $\Theta^+$ can be written as 
\begin{eqnarray*}
\mu_N &=& g_{\omega N} \omega_0 + \sqrt{k_F^2 + {m_N^*}^2},\\
\mu_{\Theta^+} &=& g_{\omega \Theta^+}\omega_0 + 
\sqrt{k_{\Theta^+}^2+{m_{\Theta^+}^*}^2},
\end{eqnarray*}
where $g_{\omega N}=3g_\omega^u$ and $g_{\omega \Theta^+} = 4 g_\omega^u$
are used following the quark counting rule. 
Here we assume $k_{\thetaplus} = 0$ and consider only a 
stationary $\thetaplus$
particle for simplicity.

%%%%%%%%%%%%%%%%%%%%%%%%%%% Fig.4  %%%%%%%%%%%%%%%%%%%%%%%%%%%%%%%%%%%%%
\begin{figure}
\begin{center}
\epsfig{file=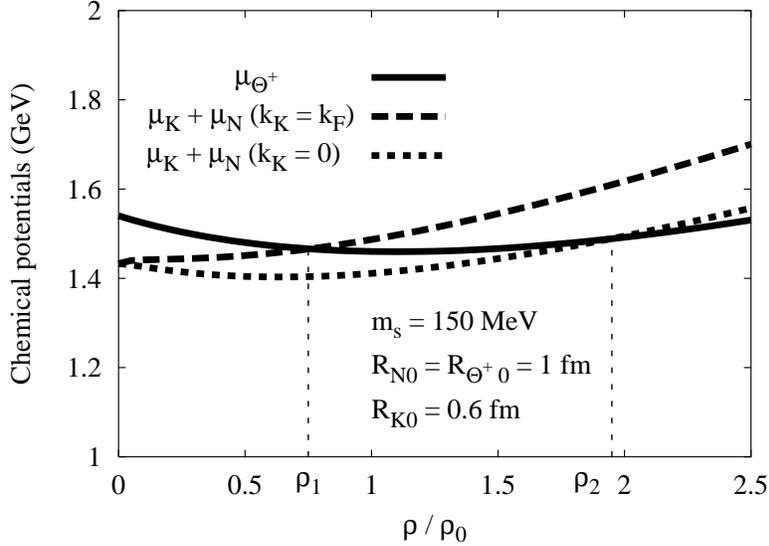,height=7.5cm,angle=0}
\end{center}
\caption{Comparison between $\mu_{\Theta^+}$ and $\mu_K + \mu_N $ in the MQMC model.
At $\rho < \rho_1$, $\mu_{\Theta^+}$ is greater than $\mu_K + \mu_N$ with
$k_K = k_F$, so both $\Theta^+ \rightarrow KN$ and
$N\Theta^+ \rightarrow NNK$ decays are possible.
At $\rho_1 < \rho < \rho_2$, only $N\Theta^+ \rightarrow NNK$ channel
is open.
At $\rho_2 < \rho$, $\Theta^+$ decay channels to $NK$ are closed.
}
\label{fig:chemical}
\end{figure}
%%%%%%%%%%%%%%%%%%%%%%%%%%%%%%%%%%%%%%%%%%%%%%%%%%%%%%%%%%%%%%%%%%%%%%%%
Fig. \ref{fig:chemical} shows that $\mu_{\Theta^+}$ (the solid curve) 
decreases with density at low densities due to 
the decrease of $\mtheta^*$ as shown in Fig. \ref{fig:qmc}.
At low densities both $\sigma$- and $\omega$-meson fields 
are approximately proportional to the density, and
competition between attraction and repulsion produces
the nuclear saturation. 
At high enough densities, $\sigma$-meson fields are
roughly proportional to $\rho^{2/3}$, 
while $\omega$-meson fields are still proportional to $\rho$. 
Thus as the density increases, repulsion becomes stronger than
attraction. 
As a result, $\mu_{\Theta^+}$ eventually increases at high densities 
as shown in Fig.~\ref{fig:chemical}.
For the chemical potential of $KN$ system, $\mu_N + \mu_K$, 
we show
two cases; one for $k_K$ = 0 and the other for 
$k_K$ = $k_F$ where $k_F$ is the Fermi
momentum of nucleons. 
$k_K=0$ gives us the minimum value of
$\mu_N + \mu_K$ 
%for a $KN$-state in nuclear medium
plotted by the dotted curve.
%The momentum conservation in the
%decay of $\Theta^+$ in the ground state ($k_{\Theta^+}$ = 0)
%does not allow
%$k_K = 0$ since the outgoing kaon 
%should have momentum equal and opposite to that of the nucleon. 
$\mu_N + \mu_K$ with $k_K = k_F$ is plotted by the dashed curve.
In the decay of a single $\Theta^+$ to $KN$, 
a kaon should have momentum at least equal to $k_F$ of nucleons
due to Pauli blocking and momentum conservation.
However, if we consider the decay of a $\Theta^+$ through 
$N\Theta^+ \rightarrow KNN$,
a kaon can have zero momentum without violating the momentum
conservation.

Let us denote the density at the intersection between
the solid curve and the dashed (dotted) curve
by $\rho_1$ ($\rho_2$): $\rho_1 \approx 0.75 \rho_0$ and 
$\rho_2 \approx 1.95 \rho_0$.
At low densities where $\rho < \rho_1$, 
$\mu_{\Theta^+}$ is greater than $\mu_K + \mu_N$ with $k_K = k_F$, and thus
both $\Theta^+ \rightarrow KN$ and $N\Theta^+ \rightarrow KNN$ 
channels are opened.
%Thus, $\Theta^+$ at low density can decay into both channels.
At $\rho_1 < \rho < \rho_2$, $\Theta^+ \rightarrow KN$ channel 
is forbidden due to Pauli blocking,
but $N\Theta^+ \rightarrow KNN$ channel is open.
Thus, if a $\Theta^+$ is created inside a heavy nucleus 
with normal nuclear density,
a $\Theta^+$ can only decay through $N\Theta^+ \rightarrow KNN$.
But if a $\Theta^+$ is created at the surface of a nucleus 
where $\rho < \rho_1$, $\Theta^+ \rightarrow KN$ may occur.
At higher densities where $\rho > \rho_2$, 
a $\Theta^+$ may be found to be stable
because both $KN$ and $KNN$ decay channels are forbidden. 
However, if a $\Theta^+$ has non-zero momentum, 
$\mu_{\Theta^+}$ increases in the form of
$\sqrt{k_{\Theta^+}^2+{m_{\Theta^+}^*}^2}$ and the 
$\mu_{\Theta^+}$ curve will move upward. 
In that case, 
both $\rho_1$ and $\rho_2$ will increase, 
and there will be significant changes in the decay scheme.
Stability of a $\Theta^+$ depends
not only on the various interactions but also
on the kinematic conditions.

\section{Summary \label{sec:summary}}

In this work we have considered the 
in-medium mass of the pentaquark baryon $\thetaplus$ 
and its decay in matter.
We have employed the quark-meson coupling model and 
the modified quark-meson coupling model
for the description of nuclear matter.
As usual, the bag model parameters 
$B_N, ~Z_N, ~B_{\Theta^+}, ~Z_{\Theta^+}, ~B_K$ and $Z_K$
are fixed to reproduce the masses of $N$, $\Theta^+$ and $K$, respectively, 
and other QMC model parameters ($g_\sigma^u$, $g_\omega^u$ and $g_\sigma^N$)
are determined from the nuclear saturation properties.
Coupling constants related to the $\omega$-meson $g_{\omega N}$, 
$g_{\omega K}$, and $g_{\omega \Theta^+}$ are 
fixed by the quark counting rules.
We treat the bag radius ($R_0$),
the valence quark mass of the $s$-quark
($m_s$), and the $\sigma$-meson -- 
$s$-quark coupling constant ($g_\sigma^s$) as free parameters.
The effective mass of a $\thetaplus$ remains more or less constant
against the variation of $R_0$ and $m_s$, 
but it changes considerably with $g^s_\sigma$. 
The mass shift is about
10\% or more at the saturation density, and 
the amount of mass reduction varies widely 
depending on the models.
%The values of uncertain parameters need to be further
%constrained in a narrow zone.

The effective mass of a kaon as a meson bag is 
compared with that as a point particle kaon.
They are in good agreement with each other if the real part of
the optical potential of antikaon, $U_{\bar K}(\rho_0)$,
is chosen as $-100$ MeV and if usual quark model
coupling constants ($g_\sigma^s = 0$) are used.
In this case, we find $g_{\sigma K} = 1.997$.
The chemical potentials of the $KN$-system and 
$\Theta^+$ are calculated to discuss
the stability of a $\Theta^+$ in medium. 
Our results indicate that a stable state of a $\Theta^+$ in 
nuclear medium is not excluded, but the stability is sensitive
to the coupling constants
($g_\sigma^s, ~g_{\sigma K}, ~g_{\omega K}, ~g_{\omega \Theta^+}$) 
and the momenta of particles in the initial and final states.
In this work, we have considered only $KN$ channels as
decay modes, but other possible modes have to be
included for a better understanding.

\section*{ACKNOWLEDGMENTS}
This work is supported by Korea Research Foundation Grant(KRF-2002-042-C00014).

\end{document}